# Unveiling the Puzzle of Brittleness in Single Crystal Iridium

Xiandong Xu[a]†, *, Sergey V. Erohin[b]†, Konstantin V. Larionov[b]†, Q. Cheng[a], Bin Gan[c], Albert Nasibulin[b], Pavel B. Sorokin[b]*

[a] College of Materials Science and Engineering; Hunan University, Changsha 410082, China.

[b] Skolkovo Institute of Science and Technology, Nobel Str. 3, Moscow, Russian Federation.

[c] Suzhou Laboratory; Suzhou, Jiangsu 215123, China.

*Corresponding author. Email: pbsorokin@misis.ru (P.B.S.); xiandongxu@hnu.edu.cn (X.D.X.)

†These authors contributed equally to this work.

**Abstract**

Materials for extreme environments require high strength yet ductile to tolerate catastrophic damage. Face-centered cubic (FCC) metals are typically ductile under stress, but single-crystal FCC iridium exhibits intrinsically brittle, limiting its wider applications. Great efforts on theoretical studies have attributed this to non-planar dislocation cores or impurities, while direct experimental evidence has remained elusive. Here we report that high-density, sessile Frank dislocation loops with zero-net Burgers vectors are the primary cause of the brittleness, identified through atomic-resolution scanning transmission electron microscopy. Through first-principles calculations, supported by discrete dislocation dynamics simulations, we reveal that these loops form via an energetically favorable transformation from mixed perfect dislocations under stress, a process unique to iridium among other FCC metals. The immobile loops act as potent barriers, drastically increasing yield strength and work hardening by impeding dislocation glide and consuming mobile dislocations. These decisive results not only deepen the understanding of the iridium brittleness, but also describe the existence of a new embrittlement mechanism inherent to the FCC lattice and not previously described in the literature. The latter may enable novel routes for property tuning across a broad class of materials, which is of paramount importance to metallurgical technology

**Significance**

Iridium is indispensable for extreme environments in aerospace and clean energy systems, yet its perplexing brittleness, a long-standing puzzle in metallurgy, severely restricts its wider use. Through atomic-scale imaging and multiscale simulations, we reveal that high-density, immobile Frank dislocation loops form under stress and act as potent barriers to dislocation motion, leading to extreme work hardening and brittle failure. This mechanism resolves a long-standing controversy in metallurgy and provides a foundation for designing more ductile iridium-based materials for extreme environments such as aerospace, nuclear systems, and high-temperature engineering.

## Introduction

Ductility is a crucial mechanical property in engineering and manufacturing, influencing a material's suitability for processes such as cold working and its ability to absorb mechanical overload. Face-centered cubic (FCC) metals are typically ductile due to their closely packed crystal structure, abundant slip systems, and low lattice friction. However, FCC iridium exhibits unusual intrinsic brittleness, leading to brittle transgranular cleavage under tension below 500 °C. [1–10] Iridium is used in various applications, including encapsulation shells in radioisotope thermoelectric generators, rocket engine coatings, and high-temperature crucibles. [11,12] These applications benefit from its high shear modulus, superior thermal stability, and excellent resistance to corrosion and oxidation. [13–15] However, the brittleness of iridium poses significant challenges for its processing and restricts its damage tolerance, limiting its broader usability in industry.

The nature of iridium's brittleness remains controversial. Some studies attribute it to external factors such as impurities, [6,16–18] while others argue that it is an inherent characteristic of the crystal. [12,19–21] Experimental observations have shown that iridium has a high stacking fault energy (SFE) of up to 420 mJ/m$^2$ and features a compact screw dislocation core as small as 0.8 ± 0.2 nm through high-resolution transmission electron microscopy (HRTEM). [22] This planar screw core can constrict to 0.4 nm [12] and subsequently transform to a non-planar core under minor shear stress of 20 MPa, with a low lattice friction of only 50 MPa. [14,23] The non-planar core is metastable and can revert to the Shockley partials on the cross-slip plane spontaneously, illustrating a stress-driven athermal cross-slip process. This mechanism was proposed to cause the brittle cleavage in iridium by promoting rapid dislocation multiplication, accumulation, and pronounced strain hardening. [2] However, this hypothesis presents several inconsistencies. Firstly, iridium exhibits standard plastic behavior before fracturing, with a strain hardening rate following the typical $\theta_{II}=F^2 d\sigma/d\varepsilon \approx \alpha\mu$ ($F$, $\mu$, and $\alpha$ is the Schmid factor, the shear modulus, and a constant, respectively) relationship observed in FCC metals. [24] Secondly, FCC Al, [25] which also features two types of screw dislocation cores - narrowly dissociated and non-dissociated - allows rapid dislocation cross-slip and multiplication like iridium but does not exhibit brittle fracture. Thirdly, despite previous HRTEM studies have identified a relatively compact screw core, [4,6,10,22] it is unable to resolve the in-plane displacements, stacking fault and core dissociation due to dynamical scattering effects. Therefore, to reveal the underlying mechanism of iridium's brittleness, a quantitative correlation between dislocation structure and its effect on mechanical performance should be established at the atomic scale.

Here we present atomic-scale experimental evidence of the dislocation structure and its correlation to iridium's brittleness using high-resolution scanning transmission electron microscopy (STEM) and theoretical calculations. Through atomic-scale imaging to characterize the dislocation core in pre-deformed iridium using high-resolution high-angle annular dark-field STEM (HAADF-STEM), we observed a high density of novel dislocation cores with zero-net Burgers vectors. These dislocation cores are identified as Frank dislocation loops, which exhibits large separation distances and high lattice distortion. Then, we propose a mechanism for the formation of Frank loops from mixed perfect dislocations, a transformation that offers greater energy benefits in iridium compared to other FCC metals. Discrete dislocation dynamics simulations were performed to assist the interpretation of the formation of Frank loops and their effect on the mechanical performance of iridium. We demonstrate that these sessile Frank loops impart high resistance to plastic deformation, thereby resulting in exceptional work hardening and inducing embrittlement in iridium.

**Experimental observations of the deformation microstructure**

The experimental single-crystal iridium rod was synthesized by electron beam floating zone melting (refer to Materials and Methods for details). The electron backscattered diffraction-inverse pole figure and X-ray diffraction pattern reveal the <111> orientation of the prepared single crystal.

A more detailed HAADF-STEM and the corresponding energy dispersive X-ray spectroscopy (EDS) map demonstrate a homogeneous distribution of iridium at atomic scale. Additionally, the EDS spectrum recorded from the HAADF-STEM image confirms the high purity of the metal, with a minor Si-$\alpha$ peak originating from the posts of Si grids used for attaching and milling TEM sample (see *SI Appendix*, Fig. S1).

To investigate the dislocation structures, the single-crystal iridium was deformed under compression to a small strain of 3%. The corresponding compressive stress-strain curve (Fig. 1A) indicates that the yield strength of the single-crystal iridium is 98 MPa. The deformation microstructures of the pre-deformed iridium were examined by STEM, as shown in Fig. 1B. The dislocations exhibit a wave-like slip pattern with pronounced interactions between them, indicative of extensive cross-slip during deformation of iridium. A closer look via the magnified STEM image in Fig. 1C reveals nanoscale dislocation loops with an average diameter of 4.33 ± 1.06 nm, as supported by line profile analysis across a representative loop (Fig. 1C, inset) and statistical measurements (*SI Appendix*, Fig. S2). These loops blend subtly into the matrix, making them difficult to detect and often overlooked.

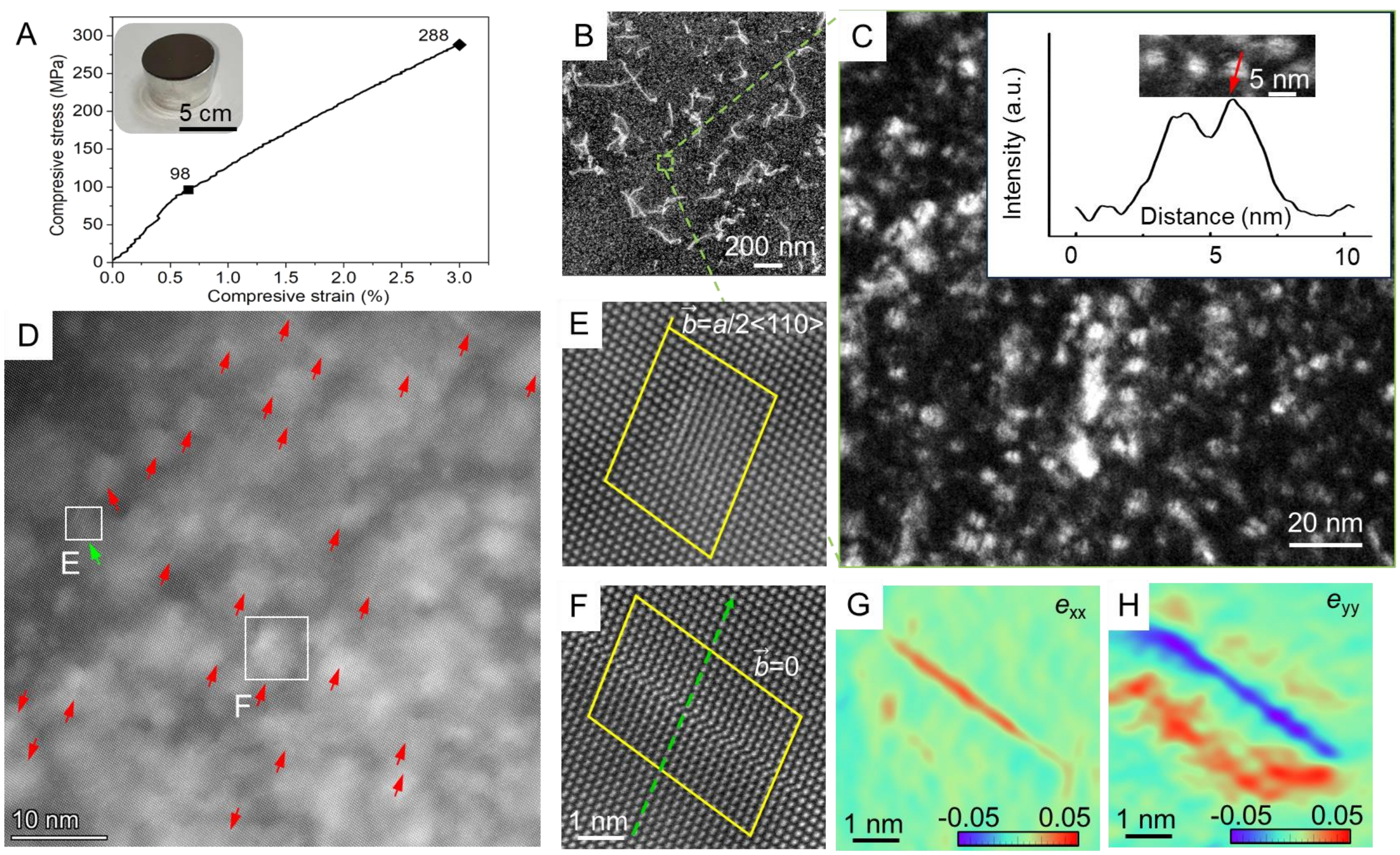

Fig. 1 The **dislocation structures of iridium at a compressive strain of 3%.** (A) Compressive stress-strain curve and a lightly deformed sample. The sample was compressed to 3% strain and unloaded. An optical micrograph of the deformed sample is shown in the top right corner. (B) A low-magnification HAADF-STEM image displaying high-density curved dislocation lines in the deformed samples. (C) A magnified HAADF-STEM image revealing nanoscale dislocation loops subtly blended into the matrix. In the inset the line profile analysis of a representative dislocation loop across red arrow, yielding an average diameter of 4.33 ± 1.06 nm is presented. (D) Atomic-resolution HAADF-STEM images of highly distorted dislocation cores. Two different dislocation cores are shown by green and red arrows. The former has b = *a*/2<110> corresponding to typical 60º mixed dislocation (E). The latter exhibits b = 0 similar to screw dislocation (F). The mean width of the zero-net Burgers vector dislocation cores is about 3.82 ± 1.02 nm. (G and H) $e_{xx}$ and $e_{yy}$ strain field maps of the loops obtained by geometric phase analysis from Fig. 1F.

Atomic-resolution HAADF-STEM imaging provides detailed information of these lattice defects (Fig. 1D), where two core structures are observed. One has a Burgers vector of $\frac{1}{2}a[110]$, corresponding to typical 60º mixed dislocation (Fig. 1E), *a* is the lattice constant. Another one exhibits a zero value of Burgers vector (Fig. 1F), similar to screw dislocation. Such curved dislocations are typical in other ductile FCC metals like Al, [26] suggesting that their presence may not substantially influence iridium's brittleness. In contrast to the low density of the 60º mixed dislocation cores (~$3.1\times10^{14}$ $m^{-2}$), the zero-net Burgers vector dislocation cores exhibit a significantly higher density of ~$7.8\times10^{15}$ $m^{-2}$ (Fig. 1D). Additionally, unlike the small rotational displacements within screw dislocation cores, the zero-net Burgers vector dislocation cores exhibit highly distorted atom columns, deviating from the perfect crystal lattice across two conjugate {111} planes, a phenomenon never observed in other highly-purity FCC metals. Based on Fig. 1F, we performed geometric phase analysis and found that strain field of dislocation cores is structured in a butterfly-like pattern. It is characterized by significant strain not only within the core regions but also on both sides of these dislocation cores (Fig. 1G and H).

From the atomic-resolution HAADF-STEM images (Fig. 1D,F) zero-net Burgers vector dislocation cores demonstrate an average separation distance of 3.82 ± 1.02 nm. This value significantly exceeds the width of observed 60º dislocation cores (1.4 ± 0.3 nm, *SI Appendix*, Fig. S3). The anomalous size of these zero-net Burgers vector dislocation cores suggests that they do not originate from screw dislocation separation. Instead, their width aligns closely with the dimensions of the dislocation loops observed in Fig. 1C (4.33 ± 1.06 nm), leading to the hypothesis that these highly distorted cores represent a unique structural manifestation of the dislocation loops.

Finally, we validate that observed unique dislocation cores is some intrinsic attribute of iridium crystal rather than external impact. First, uniform elemental distribution at the atomic scale in the STEM-EDS mapping (*SI Appendix*, Fig. S4) demonstrates that there are no impurity and chemical segregations in the defects, confirming that the strain fields solely result from the local displacement of iridium atoms. Secondly, the exceptional surface quality of the TEM sample after 2 kV final polishing with Ga ion beam and the stability of the FCC lattice of iridium under electron beam illumination for several minutes suggesting that Ga ion and electron beam exposure have minor impacts on the structure of iridium (*SI Appendix,* Fig. S5). Next, the zero-net Burgers vector dislocation cores were subjected to the electron beam for several minutes. The dislocation cores demonstrate high stability, confirming that these structures are induced by deformation rather than beam effects.

**Formation mechanism of the Frank loops**

To identify the type of dislocations observed in Ir, several dislocation models characteristic of the FCC lattice were constructed. Among these, the model that showed the closest agreement with the experimental image – in terms of local atomic strain and a zero net Burgers vector – was the Frank loop model (see Fig. 2A).

Frank loops are formed by the aggregation of vacancies or interstitial atoms [27]. Nevertheless, it remains unclear which type of Frank loop – interstitial or vacancy – is responsible for the observed experimental phenomenon. To directly distinguish between vacancy and interstitial loops, both models were constructed. For interstitial loop, a finite-size section of the atomic {111} plane was inserted into the iridium crystal, as shown in Fig. 2B. For a vacancy a part of the atomic {111} plane was deleted.

Then the strain distribution of interstitial and vacancy-type Frank loops from the theoretical atomic images under [011] zone axis was analyzed. The interstitial-type loops exhibit a tensile strain component at the dislocation core, whereas vacancy-type loops have dislocation cores displaying compressive strain (Fig. 2C and *SI Appendix*, Fig. S6). The dislocation cores observed in experiments show tensile strain (Fig. 1G), indicating that they are likely interstitial-type loops. Although the integrated HAADF-STEM intensities across the loops along the (111) plane display some variation between the idealized atomic model and actual experimental samples (Fig. 2D) due to thickness disparities, the consistent patterns in intensity distribution provide strong evidence that the observed loops are interstitial-type.

The results unequivocally indicate that the observed dislocations are interstitial Frank loops. This, however, raises the question of their formation, given that these dislocations are typically a consequence of irradiation. Moreover there are some evidence that Frank loops often are not stable and transforms to prismatic loops [28].

The main difference between the Frank and prismatic type lies in the shift of the inserted atoms (*SI Appendix*, Fig. S7A and B, respectively), where Frank loops are typically associated with stacking faults, whereas prismatic loops are not. After optimization using a pre-trained machine learning interatomic potential (MLIP, Methods, *SI Appendix*, Fig. S8), the resulting prismatic loop does not match the experimentally observed distortions in the dislocation core region, as shown by the comparison between Fig. 1F and *SI Appendix*, Fig. S7C.

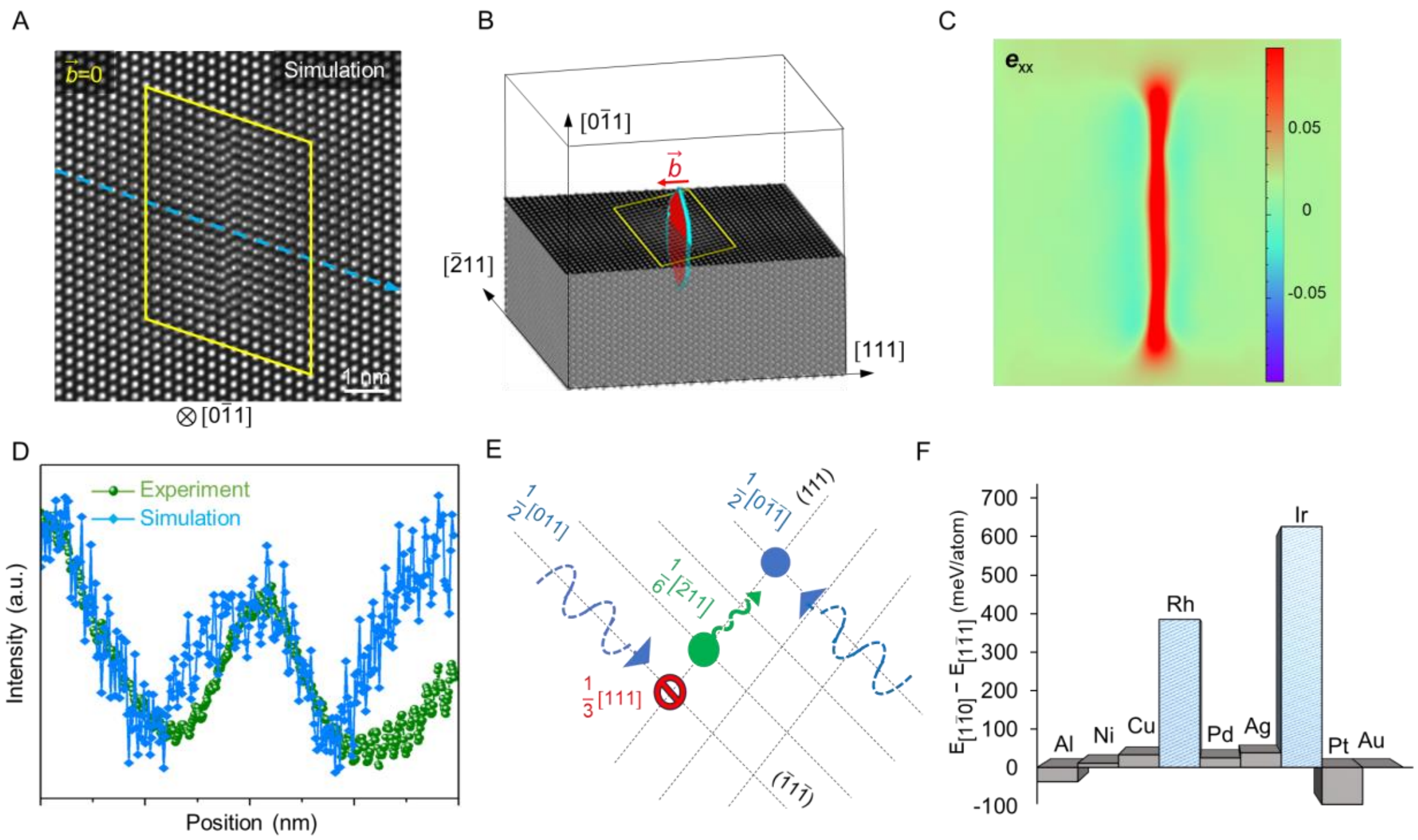

Fig. 2 **Dislocation model and possible reaction mechanism.** (A) Simulated HAADF-STEM image based on the atomic model. For direct comparison with the experiment (Fig. 1F), the atomic structure oriented by the [0$\bar{1}$1] direction is shown. The yellow contour represents the zero-net Burgers vector. (B) Perspective view of the iridium crystal structure atomic model with a Frank loop (colored by cyan) formed by an inserted atomic plane (colored by red). (C) Theoretical $e_{xx}$ map of an interstitial-type loop corresponded to the atomic structure (A). (D) Integrated HAADF-STEM intensities from the simulated interstitial-type loops in (A) across blue dashed arrow and from the experimental green dashed arrow in Fig. 1F. (E) Proposed reaction of a mixed perfect dislocation pair with a Shockley partial dislocation, illustrating the formation of the Frank loop.

Parallel dashed lines indicate the (111) and $(\bar{1}1\bar{1})$ plane systems. Wavy arrows show the direction of glissile dislocation movement, with blue and green circles representing mixed perfect and Shockley dislocations, respectively. The red circle denotes the position of the Frank partial. (F) DFT calculated energy differences between the two types of dislocations per embedded atom. Positive and negative values indicate energetically favorable formation of two Frank partials and two mixed perfect dislocations, respectively.

The formation of the high-density dislocation loops in single-crystal iridium cannot be explained by diffusion mechanisms alone, as the required concentration of interstitial atoms would be improbable. Instead, we propose a transformation mechanism where Frank loops form from mixed perfect dislocations (containing both edge and screw components). This hypothesis is supported experimentally by the presence of Frank loop nearby mixed perfect dislocations (*SI Appendix*, Fig. S9). These mixed perfect dislocations are highly glissile and can move within the primary slip system. When a pair of dislocations with opposite signs glide along parallel $(\bar{1}1\bar{1})$ planes and approach each other closely, they can form a dislocation dipole and react. The proposed reaction can be derived through the Thompson tetrahedron (*SI Appendix*, Fig. S10):

$$DB = D\delta + \delta B\text{: } \tfrac{1}{2}a[011] \rightarrow \tfrac{1}{3}a[111] + \tfrac{1}{6}a[\bar{2}11]$$

where a mixed perfect dislocation ($\frac{1}{2}a[011]$) dissociates into a glissile Shockley partial dislocation ($\frac{1}{6}a[\bar{2}11]$, as shown by green in Fig. 2E) and a sessile Frank partial ($\frac{1}{3}a[111]$, as shown by red in Fig. 2E). After that, the resulting Shockley partial may interact with another $\frac{1}{2}a[0\bar{1}\bar{1}]$ perfect dislocation moving along a neighboring plane $(\bar{1}1\bar{1})$, leading to the formation of a second $\frac{1}{3}a[\bar{1}\bar{1}\bar{1}]$ Frank partial nearby the first one. This creates an immobile pair of Frank partials with a net-zero Burgers vector. Structurally and functionally, this pair of Frank partials is a 1D analogue of a Frank loop.

The energy gain from this dislocation reaction can be estimated by the Frank's rule, [27] where the energy of a dislocation is proportional to the square of the magnitude of its Burgers vector ($\sim\boldsymbol{b}^2$). In this scenario, the produced Frank partials are more energy favorable than the initial pair of mixed perfect dislocations. However, the movement of the Shockley partial inevitably gives rise to a stacking fault, which is associated with notably high energy. Additionally, it has been suggested that in FCC metals, the reverse transformation of a Frank loop to prismatic one is more energy favorable, particularly for large loops. [28] Given that iridium possesses the highest SFE among other FCC metals, one would expect a similar pathway. However, our Density Functional Theory (DFT) calculations revealed unexpected results.

By inserting a periodic ribbon of small size within the iridium structure (Methods and *SI Appendix*, Fig. S7D), we calculated the energy of pairs of Frank dislocations and mixed perfect dislocations and demonstrated their energy difference, with a positive value indicating the energy favorability of the former. An intriguing pattern emerges when comparing iridium with other FCC metals, as shown in Fig. 2F. The energy difference does not directly correlate with the SFE or shear modulus but rather uniquely distinguishes iridium and rhodium from other FCC metals. While the focus of this study lies on iridium, it is worth noting that there are experimental evidence suggesting that rhodium also demonstrate anomalous brittleness. [7] Thus, iridium may possess a unique mechanism for Frank loops formation. The presented DFT energy calculations reveal a practically one-way reaction favoring the formation of these immobile loops, making them exclusively stable in iridium.

**Effect of the Frank loops on mechanical properties**

Building on the atomic-scale observations of high-density Frank loops, we employed discrete dislocation dynamics (DDD) using ParaDiS [29] to directly validate the transformation pathway from perfect dislocations to Frank loops under stress. In contrast to above DFT analysis, DDD simulations can explicitly capture dislocations reactions. The mechanism from Fig. 2E may be represented using single mixed perfect dislocations with arcs as depicted in Fig. 3A. External stress initiates a Shockley partial (green), generating a sessile Frank partial (red). Continued glide of the Shockley segment and the original dislocation closes the arc, ultimately pinching off to form an isolated Frank loop. This dynamic illustrates how the experimentally observed loops can originate from stress-driven dislocation reactions.

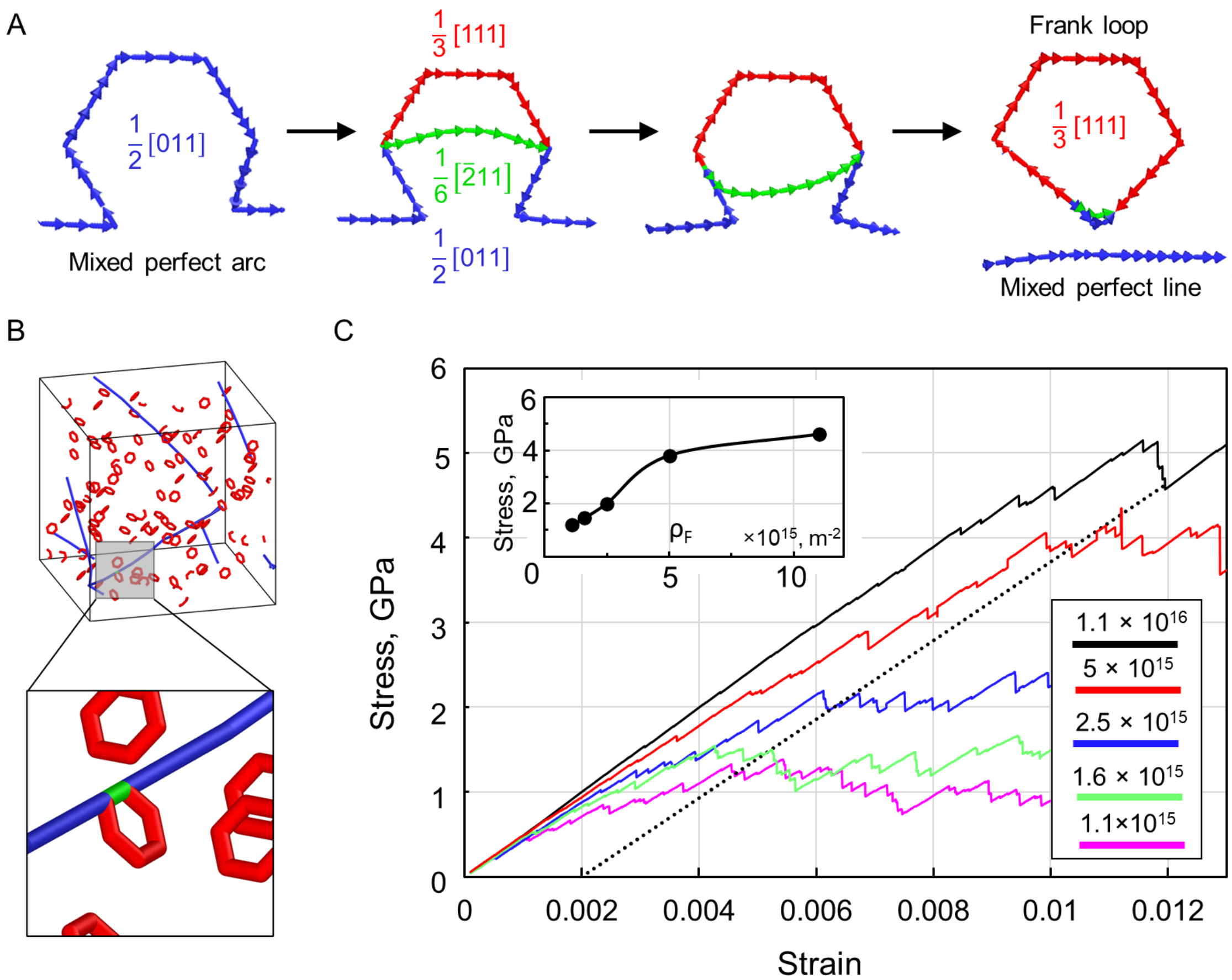

Fig. 3 **Dislocation dynamics analysis.** (A) Simulated process of Frank loop formation (red) from mixed perfect arc (blue) via Shockley partial (green) gliding. (B) Example of dislocation network for $\rho_{\mathrm{loop}} = 2.5\times10^{15}$ m$^{-2}$. Zoomed region illustrates a mixed perfect line pinned by a Frank loop. (C) Stress-strain curves for various densities of Frank loops while keeping number of mixed perfect dislocations to be constant. Yield strength dependence on the density of Frank loops at the 0.2% proof-strain is shown as the inset.

In addition to elucidating the formation mechanism, we quantified the impact of the high-density Frank loops on the mechanical properties of iridium by simulating stress-strain behavior across various loop densities ($\rho_{\mathrm{loop}}$). The initial dislocation network consisted of two primary defect types: glissile mixed perfect dislocations with $\boldsymbol{b} = \frac{1}{2}a[011]$ and sessile Frank loops with $\boldsymbol{b} = \frac{1}{3}a[111]$. An illustrative example of a dislocation configuration is presented in Fig. 3B, with the $\rho_{\mathrm{loop}}$ set at $2.5\times10^{15}$ m$^{-2}$, where zoomed region shows Frank loops pinning mobile dislocations. Large-scale simulations were performed for the generated dislocation networks, and stress-strain curves were derived, as depicted in Fig. *3*C. The concentration of glissile mixed dislocation lines remained constant at $6.3\times10^{14}$ m$^{-2}$, while the $\rho_{\mathrm{loop}}$ varied from $1.1\times10^{15}$ m$^{-2}$ to $1.1\times10^{16}$ m$^{-2}$, consistent with experimental estimations obtained from HAADF-STEM images (Fig. 1D). To mitigate the stochastic nature of dynamic results, each curve was obtained by averaging multiple runs with random initial positions and orientations of both mixed dislocation lines and Frank loops.

Detailed analysis in Fig. *3*C reveals a monotonic increase in yield strength with rising $\rho_{\mathrm{loop}}$. For the first three curves (up to $2.5\times10^{15}$ m$^{-2}$), the elastic regime remains relatively narrow, staying below the 0.5% strain level. However, starting from $5\times10^{15}$ m$^{-2}$, the elastic region extends significantly, reaching 1% strain and beyond. Furthermore, to further analyze the strain hardening effect, 0.2% proof yield strength was extracted from the stress-strain curves, as summarized in the inset of Fig. *3*C. Starting from a minimum value of approximately 1.2 GPa for the lowest $\rho_{\mathrm{loop}}$ studied, the yield strength undergoes a rapid several-fold increase, eventually approaching a plateau of 4.6 GPa. Notably, the yield strength increases by 3.8 times with each order of magnitude rise in $\rho_{\mathrm{loop}}$. These observations suggest that the high density of sessile Frank loops acts as strong obstacles to mobile dislocations, effectively suppressing dislocation-mediated plastic flow and inducing significant stress localization. This pronounced inhibition of plastic relaxation results in accelerated work hardening and ultimately triggers brittle fracture, as the accumulated strain energy cannot be adequately dissipated through conventional ductile deformation mechanisms.

**Summary remarks**

In conclusion, this work provides a significant leap in understanding the intrinsic brittleness of high-purity single-crystal iridium. Through a multiple approach, we found that its brittleness is associated with the formation of high-density compact sessile Frank loops. These sessile defects, identified through atomic-resolution microscopy, exhibit unique structural characteristics such as zero-net Burgers vectors and pronounced lattice distortions that distinguish them from conventional dislocation cores. The embrittling action operates through dual mechanism: (1) generation of high-density immobile loops from glissile dislocations via reaction cascades; and (2) dynamic depletion coming from geometric constraint due to non-planar core distortions that generate localized strain fields, raising Peierls stresses and progressively paralyzing plastic flow.

The anomalous high concentration of Frank loops is proposed to originate from mixed perfect dislocations as verified by DFT from substantial energy gain in reaction. Such transformation is found to be unique to iridium and potentially rhodium among other FCC metals. Our HAADF-STEM images of proximity of mixed perfect dislocations and Frank loops support the above hypothesis. Furthermore, dislocation dynamics simulation explicitly demonstrates the increased work hardening rate caused by immobile loops formation. This, in turn, produces strong resistance to a plastic deformation, consequently leading to brittle failure. The uncovering of stress-induced Frank loop dominance in iridium not only solves a decades-old puzzle in metallurgy but also paves the way for tailoring dislocation pathways to enhance toughness in extreme engineering applications. These findings open new avenues for material design: modulating Frank loop formation via alloying (e.g., W, Re additions) could mitigate brittleness while preserving iridium's extreme environmental stability.

## Materials and Methods

**Sample preparation.** The iridium single crystal was prepared in a 25 KW electron beam floating zone melting (EBFZM) furnace. [30,31] The iridium rod used for EBFZM was 200 mm in length and 8 mm in diameter. Prior to the melting, the vacuum level inside the furnace was pumped down to $1\times10^{-4}$ Pa using a molecular pump. During the melting, the single crystal iridium was grown at a pulling rate of 3 mm/min. The temperature of the molten zone was controlled by the power of the EBFZM. Initially, the power was gradually increased until the molten zone formed, and then this power was maintained to stabilize the molten zone. In this experiment, the electron beam gun was set to move at a speed of 10 mm/min, with the EBFZM power at 0.8 KW, and the height of the molten zone was approximately 5-8 mm. At this point, the mean free path of molecules was about 14 m, and the long diffusion distance ensured the stability of crystal growth during the EBFZM process.

**Mechanical tests.** Compressive deformation was performed at room temperature using a Landmark MTS servo-hydraulic test system with a strain rate of $5\times10^{-3}$ $s^{-1}$ and a strain of 3%. The cylindrical samples had dimensions (diameter × height) of 8×6 $mm^2$. The height after compression was measured and height reduction was calculated to calibrate the sample strain.

**Microstructure characterization.** The growth orientation of the single-crystal iridium was examined by XRD on a Rigaku SmartLab SE instrument. The XRD tests were conducted at 45 kV and 45 mA, with a scanning range of $2\theta$ from 30º to 100º and a scanning rate of 0.5º/min. Electron backscatter diffraction (EBSD) images were captured on an SEM (Zeiss, Gemini 300) equipped with a step size of 1.5 μm. The detail analyses of dislocation fine structures were performed using a DCI-STEM technique on an aberration-corrected TEM (Thermoscientific, Themis Z 3.2), operating at 300 kV with a resolution of about 60 pm. The probe forming convergence semi-angle of the illumination is 17.8 mrad, and the inner-to-outer collection angle of the HAADF detector ranges from 52 to 200 mrad. The strain field mappings of the dislocation cores were obtained by geometric phase analysis using the Strain++ software. High-resolution EDS mappings were obtained to study elemental distribution within dislocation cores. TEM specimens were prepared using a focused ion beam lift-out method.

HAADF-STEM image simulation of the interstitial-type Frank loop was performed using Prismatic software [32] based on the atomic model in Fig. 2B, with a specimen thickness of approximately 5 nm achieved by replicating the input crystal structure along the projection axis. The simulation parameters included an acceleration voltage of 300 kV, a convergence semi-angle of 18 mrad, a detector collection angle of 52 to 77 mrad, a scan step of 0.001 nm/pixel, a spherical aberration (C30) of 10 μm, 10 frozen phonon configurations, and an interpolation factor of 10 for the plane-wave reciprocal-space interpolated scattering matrix algorithm.

**DFT calculations.** For machine learning potential (MLP) preparation, ground-state calculations of various iridium structures are performed with the DFT method within the general gradient approximation (GGA) functional in the Perdew-Burke-Ernzerhof (PBE) parameterization. [33] We use the projector augmented wave method with the periodic boundary conditions as implemented in the Vienna Ab-initio Simulation Package. [34,35] A plane-wave energy cut-off is set to 400 eV. The Brillouin zone is sampled according to the Monkhorst-Pack schemes with a k-point spacing of 0.16 $Å^{-1}$ for the relaxation. To avoid spurious interactions the translation vector along non-periodic direction is set to be 15 Å. In addition to the ground-state, ab initio MD calculations (T = 600 K) were performed as well to increase the number of available configuration points for MLP training.

To compare the energy between Frank and mixed perfect dislocations the atomic models of metals with embedded ribbons were optimized. The structure of ribbon corresponding to a pair of Frank dislocations is represented in *SI Appendix*, Fig. S7D (its stacking fault matches *SI Appendix*, Fig. S7A). The total number of atoms is 187. The ribbon consists of 7 atoms with width of ~1 nm. For Ir case, the box size corresponds to 2.74 × 2.12 × 0.47 $nm^3$. As for mixed perfect dislocations, its periodic DFT model looks identical to the Frank one (*SI Appendix*, Fig. S7D) except the shift of a ribbon along $[\bar{2}11]$ in accordance with stacking fault from *SI Appendix*, Fig. S7B.

**MLP training and calculations.** Here we implement MLP with moment tensor descriptors to train the potential. iridium bulk structure, slabs, 1D nanowires in a defect-free state and with multiple defects were considered to generate the training set, see *SI Appendix*, Fig. S8 for detail. The resulting training set consists of 300 DFT optimized structures and corresponding ground-state energy values as well as atomic forces listed for 35762 atoms. Next, the ML potential was trained using the MLIP-2 package [36] which belongs to the moment tensor potential family. 24$^{th}$ level of potential was used as a template with the cutoff radius of 5 Å and the basis size equal to 8. Energies, forces and stresses from the training set were weighted as 1, 0.01, and 0.001, respectively. Resulting RMS absolute difference in energy and atomic forces was found to be 0.003 eV/atom and 0.15 eV/Å, respectively. Comparison of energy and forces between DFT and MLP methods is directly plotted in *SI Appendix*, Fig. S8, which demonstrates high accuracy of generated potential.

A finite-size section of the atomic {111} plane was inserted into (or deleted from) the iridium crystal to create interstitial(vacancy)-type Frank loop. An atomic model of a single interstitial-type Frank loop with a diameter of 4 nm (comprising a total of 50 thousand atoms) is represented in *SI Appendix*, Fig. S6. These models were optimized using an ML-trained potential in LAMMPS package. [37] It is important to note that visible atomic displacement is identical to that of Frank partial pair model investigated by DFT method. Moreover, our ML-trained potential confirms that Frank loop is energetically more favorable than prismatic one constructed by mixed perfect dislocations. The latter findings correlate with the results of direct DFT calculations as revealed in Fig. 2F.

**Dislocation dynamic simulations.** ParaDiS software was used to study mechanism of Frank loop formation from a mixed perfect dislocation and investigate the mechanical properties of Ir crystal. In this method, each dislocation is represented as a network of nodes connected by straight segments defined by the Burgers vector. By calculating of nodal forces through the Peach-Koehler formula, [38] employing mobility functions, and incorporating local stress fields, it is possible to track the dislocation motion for large-scale scenarios with a linear-time complexity.

Fast Multipole method was implemented for far-field interactions with the second order of multipole expansion and the fifth order of Taylor expansion. The shear modulus, Young's modulus and Poisson's ratio were set as 212 GPa, 534 GPa and 0.26, respectively. In our study, a uniaxial loading with a constant strain rate ($\dot{\varepsilon}$ = 5×$10^3$ $s^{-1}$) regime was used, allowing for cross-slipping process. The simulation box, set under periodic boundary conditions, was configured as a cube with a size of $L$ = 0.09 µm. Glide motion in the FCC system was constrained to the {111} slipping plane family. For dislocation analysis in MLP and DDD methods, OVITO software is used. [39]

## Data availability

All data needed to evaluate the conclusions in the paper are present in the paper and the Supplementary Materials.

## Author contributions

X.D.X. initiated, supervised the project and conducted the experiments. B.G. provided the sample. Q.C. a small portion of the transmission electron microscopy (TEM) experiments and drafted the initial manuscript. S.V.E. and K.V.L. conducted DFT calculations, machine learning, and dislocation dynamics simulations under P.B.S. supervision. Q.C., S.V.E., and K.V.L. drafted the original manuscript. All coauthors participated in discussions and manuscript revisions.

## Acknowledgments

This work was sponsored by the National Key Research and Development Program, Major Scientific Instrument Special Program for Basic Research - Development and Application of All-domestic Three-dimensional Atom Probe Precision Measurement Instrument Project (Grant No. 2023YFF0716200), and the Science and Technology Innovation Program of Hunan Province (Grant No. 2024RC1035). We would like to thank Professor Mingwei CHEN (Southern University of Science and Technology), Professor Zhaoxuan WU (City University of Hongkong), Professor Xiuyan LI (Chinese Academy of Sciences (CAS)-Shenyang National Laboratory for Materials Science), and En MA (Xi'An Jiaotong University) for their helpful discussions.

## Competing interests

The authors declare that they have no competing interests.

# Supporting Information

**This document includes:**

**Supplementary Figures 1 -10**

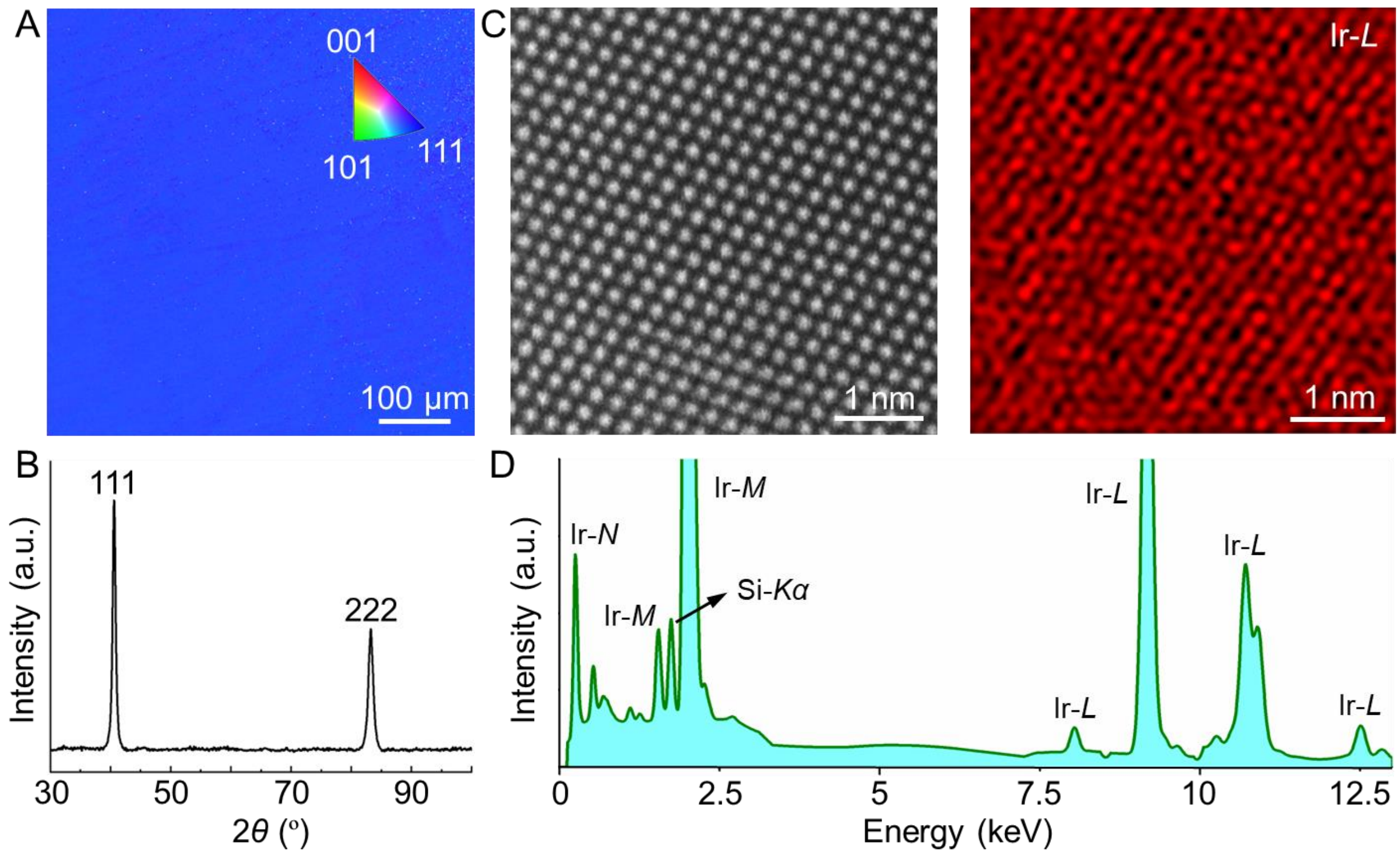

**Figure S1 The crystal structure and composition of iridium.** (A) Electron backscattered diffraction-inverse pole figure of the surface of the prepared iridium rod showing a <111> single-crystal orientation. The inset is the color-code for cubic materials. (B) XRD pattern further confirming the <111> crystal orientation. (C) A HAADF-STEM image under [110] zone axis and corresponding EDS map of Ir-*L*. (D) EDS spectra recorded from the HAADF-STEM image in C.

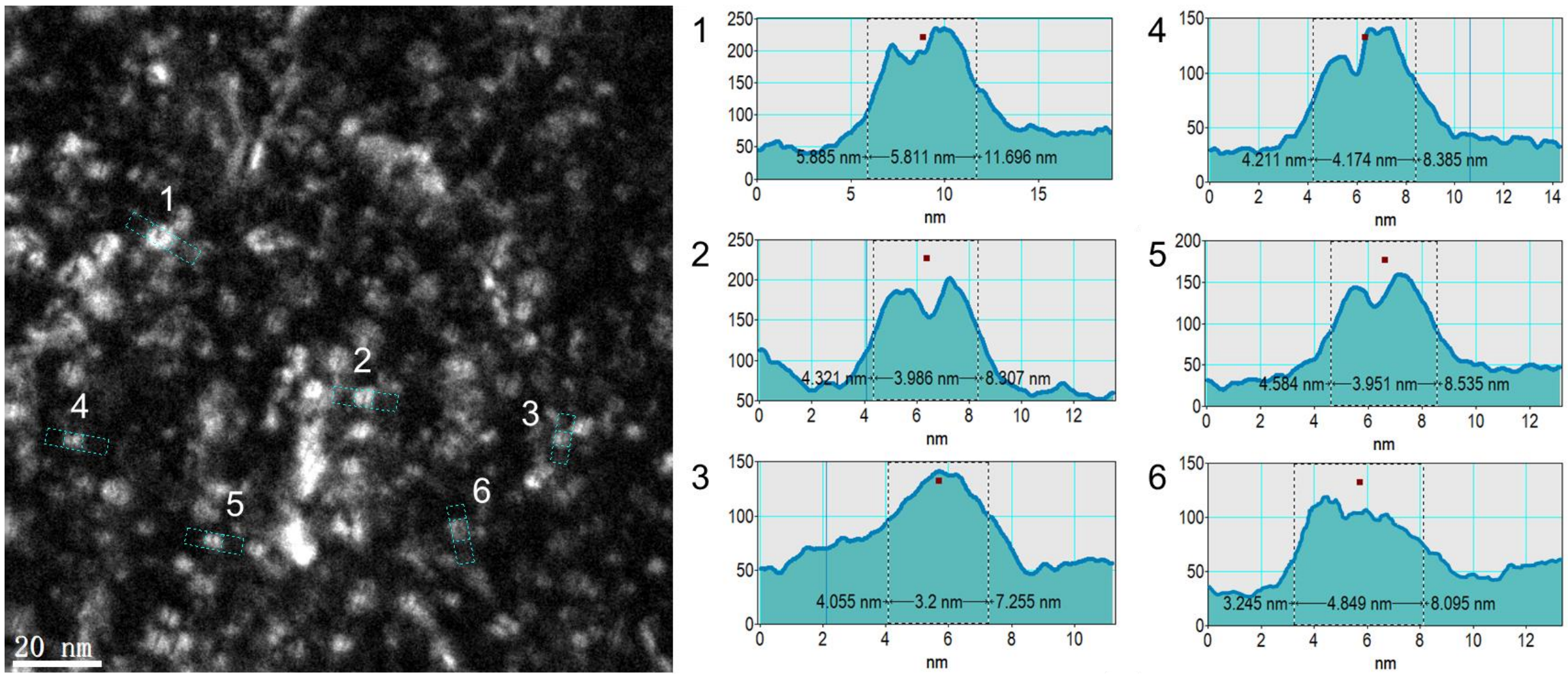

**Figure S2 Magnified HAADF-STEM image and integrated intensities along the lines across the dislocation loops 1-6**. The average diameter of the dislocation loops was estimated to be 4.33 ± 1.06 nm by the full width at half maximum of the integrated intensities along lines 1-6.

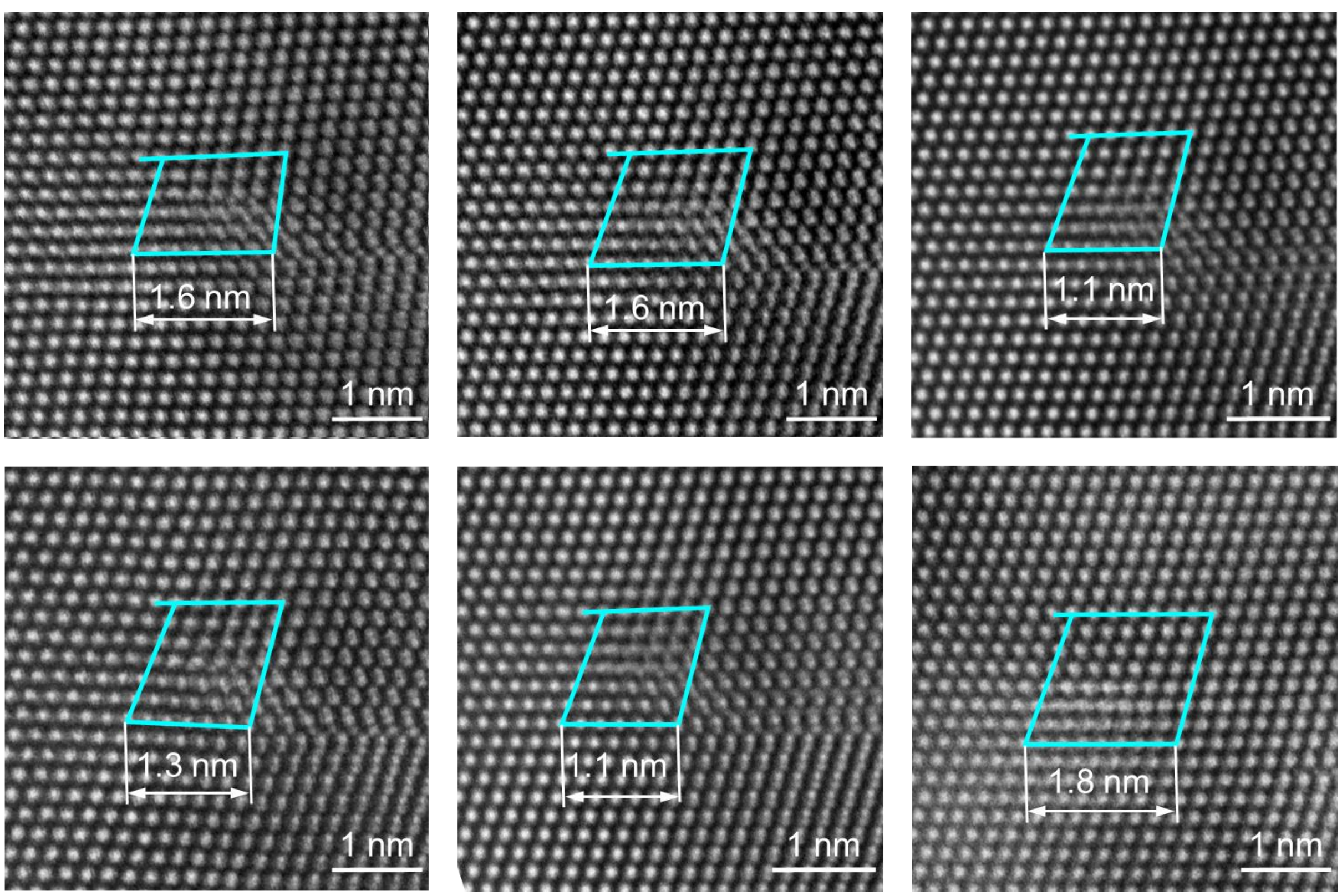

**Figure S3 HAADF-STEM images of 60º mixed dislocations.** The measured separation distance of 60º mixed dislocations is 1.4 ± 0.3 nm, corresponding to a high SFE of 352 mJ/m$^2$. This SFE calculation utilizes the equation: $\gamma = \frac{\mu b_p^2}{8\pi d}(\frac{2-\upsilon}{1-\upsilon})(1 - \frac{2\upsilon\cos(2\beta)}{2-\upsilon})$, where $\mu$ is the shear modulus (212 GPa for Ir). $b_p$ = 0.147 nm is the Burgers vector of partial dislocation, and $\upsilon$ = 0.26 is the Poisson's ratio.

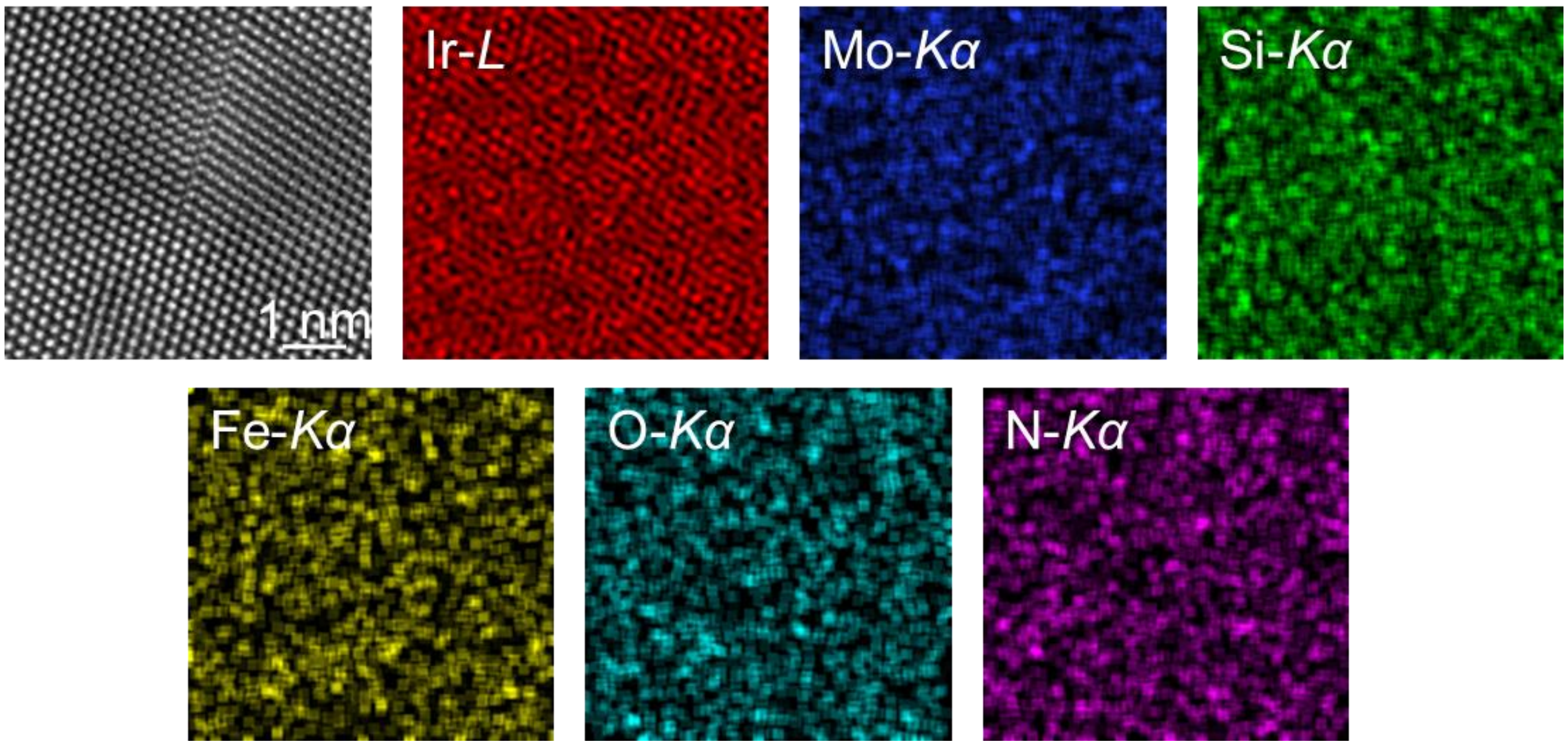

**Figure S4 HAADF-STEM image and EDS mapping of the zero-net Burgers vector dislocation core.** No elemental segregation along the zero-net Burgers vector dislocation core is observed.

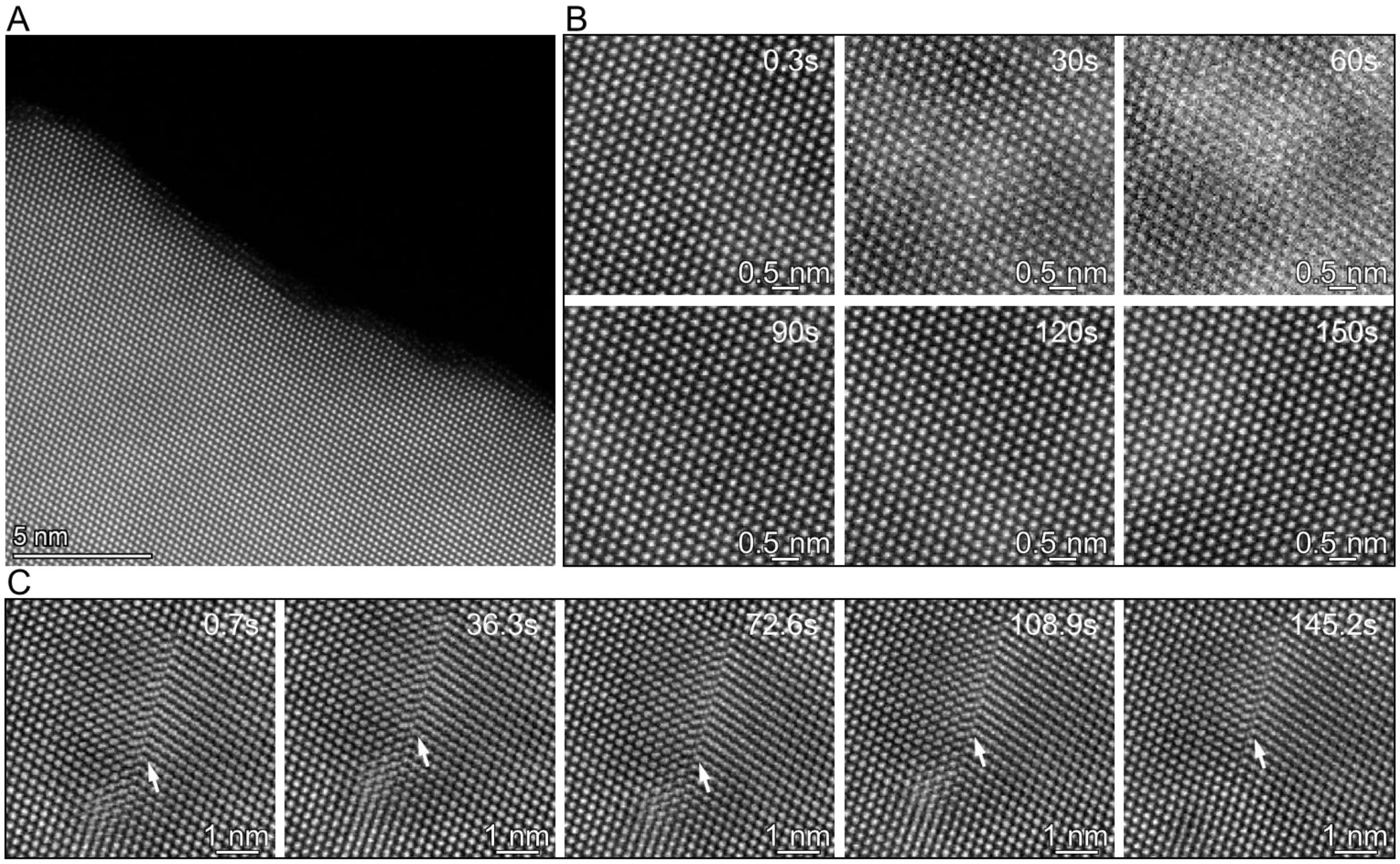

**Fig. S5 The surface quality of the TEM sample and the effects of electron beam illumination on the iridium lattice and zero-net Burgers vector dislocation cores.** (A) Following final polishing with a 2 kV Ga ion beam, the TEM sample exhibited exceptional surface quality with no apparent amorphization. (B) HAADF-STEM images of the iridium lattice, exposed to electron beam illumination for several minutes with an electron current of 38 pA, demonstrated the stability of the iridium lattice under the exposure. (C) HAADF-STEM images of the zero-net Burgers vector dislocation cores, also under electron beam illumination for several minutes with an electron current of 8 pA, revealed high stability in the dislocation cores, confirming that these structures are a result of deformation rather than beam effects.

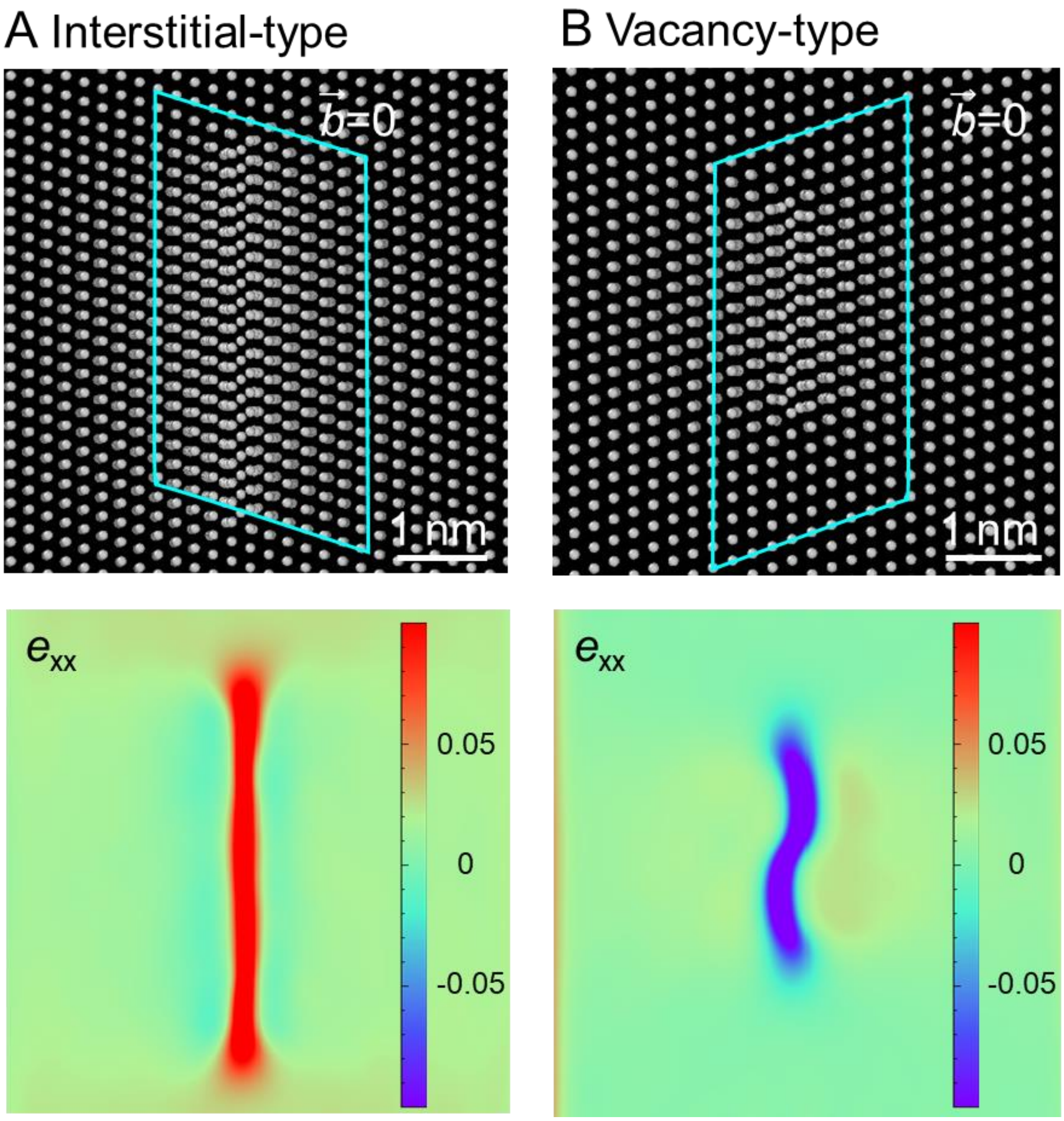

**Figure S6 Atomic images and corresponding $e_{xx}$ maps of theoretical models for two-type Frank loops.** (A) Theoretical atomic image and $e_{xx}$ map of an interstitial-type loop. (B) Theoretical atomic image and $e_{xx}$ map of a vacancy-type loop.

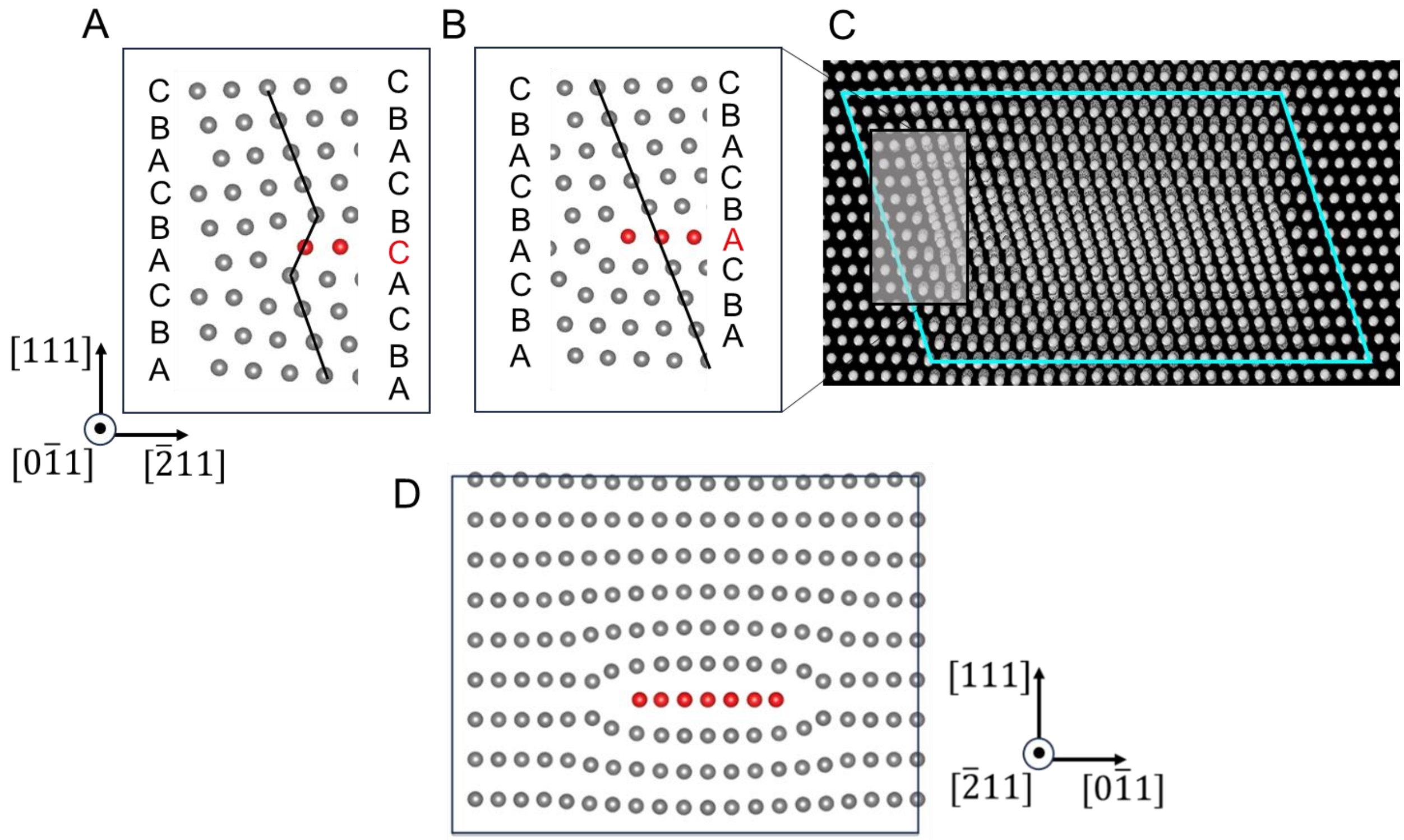

**Figure S7 Atomic models.** Slices of [0$\bar{1}$1] oriented Ir crystal with (A) Frank loop and (B) prismatic loop. Two stacking types of {111} planes are shown. Full atomic model of [0$\bar{1}$1] oriented Ir crystal with prismatic loop is shown in (C). Yellow contour depicts zero net Burgers vector in (C). Slice from (B) is extracted from shadowed area in (C). (D) DFT atomic model of bulk Ir with inserted ribbon. The periodicity of the ribbon is along [$\bar{2}$11] direction.

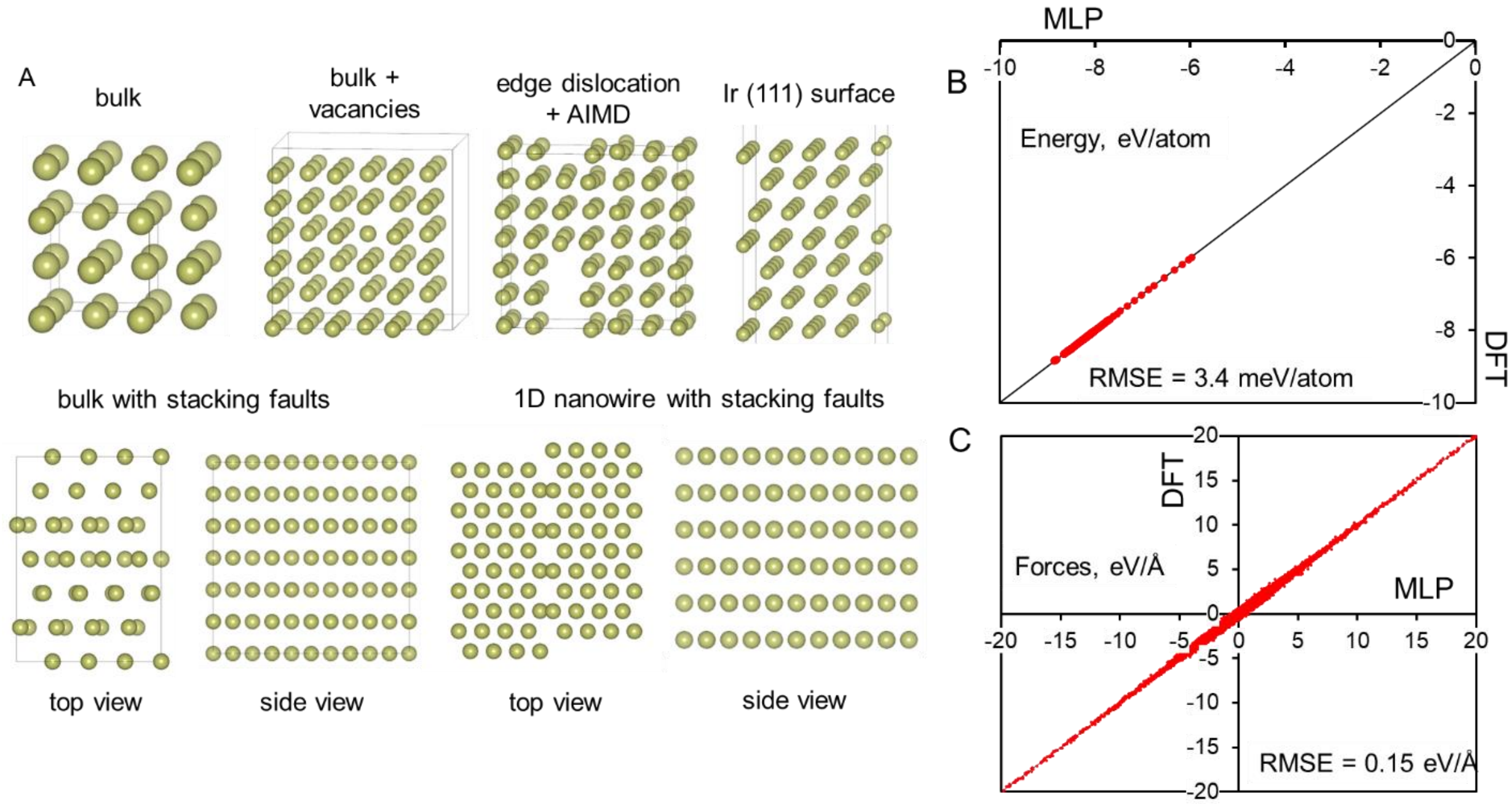

**Figure S8** (A) Examples of Ir atomic structures for MLP training set. perfect bulk, bulk with vacancies, edge dislocation via AIMD simulation, Ir(111) surface, bulk with stacking faults and 1D nanowire with stacking faults are demonstrated.

Comparisons of energy (B) and atomic forces (C) between MLP and DFT methods.

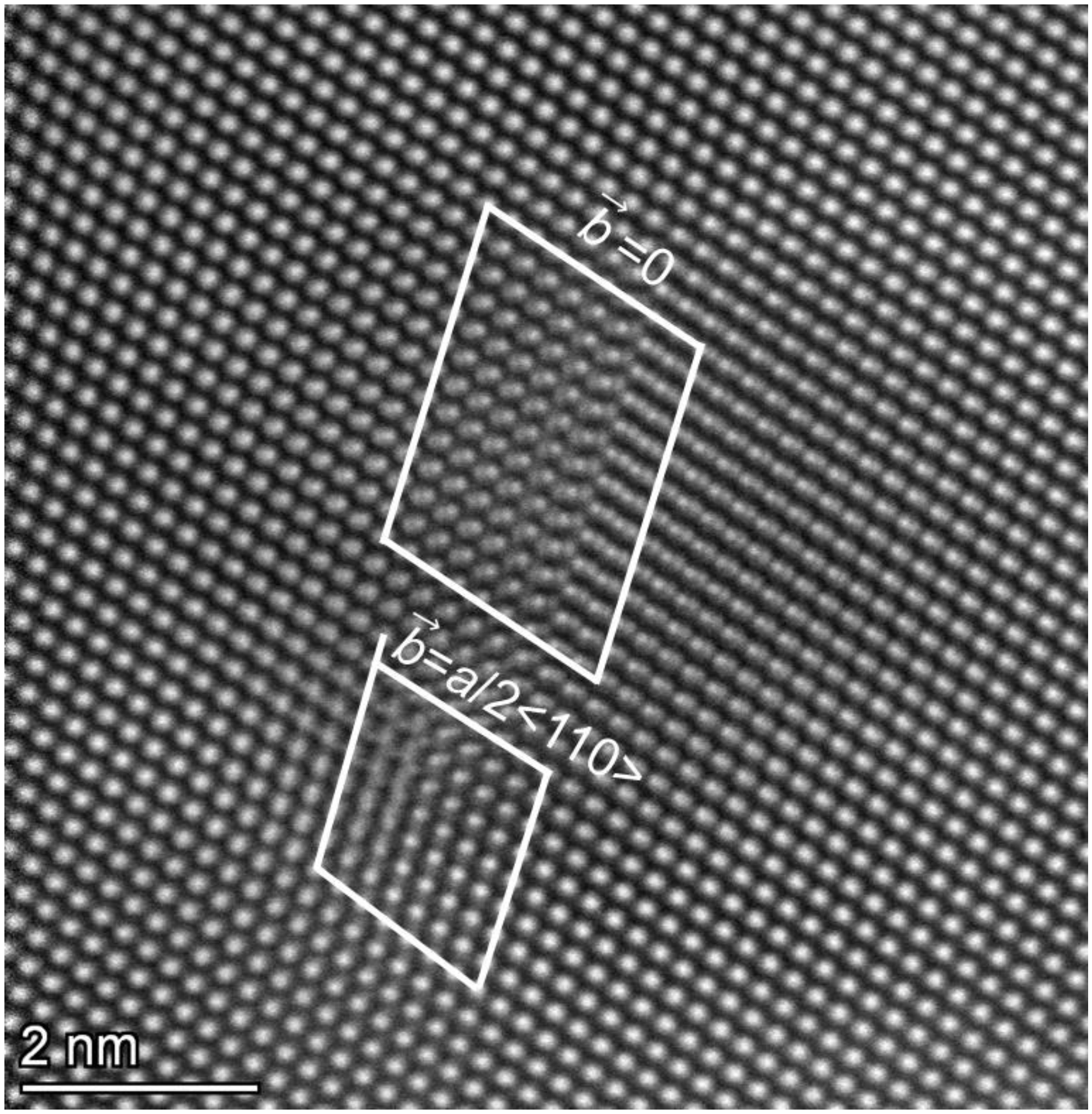

**Figure S9 High-magnification HAADF-STEM images of dislocation cores in compressive iridium.** Two different dislocation cores are observed. One has a Burgers vector of 1/2<110>, corresponding to typical 60º mixed dislocation. The other one exhibits a zero value of Burger vector, similar to screw dislocation.

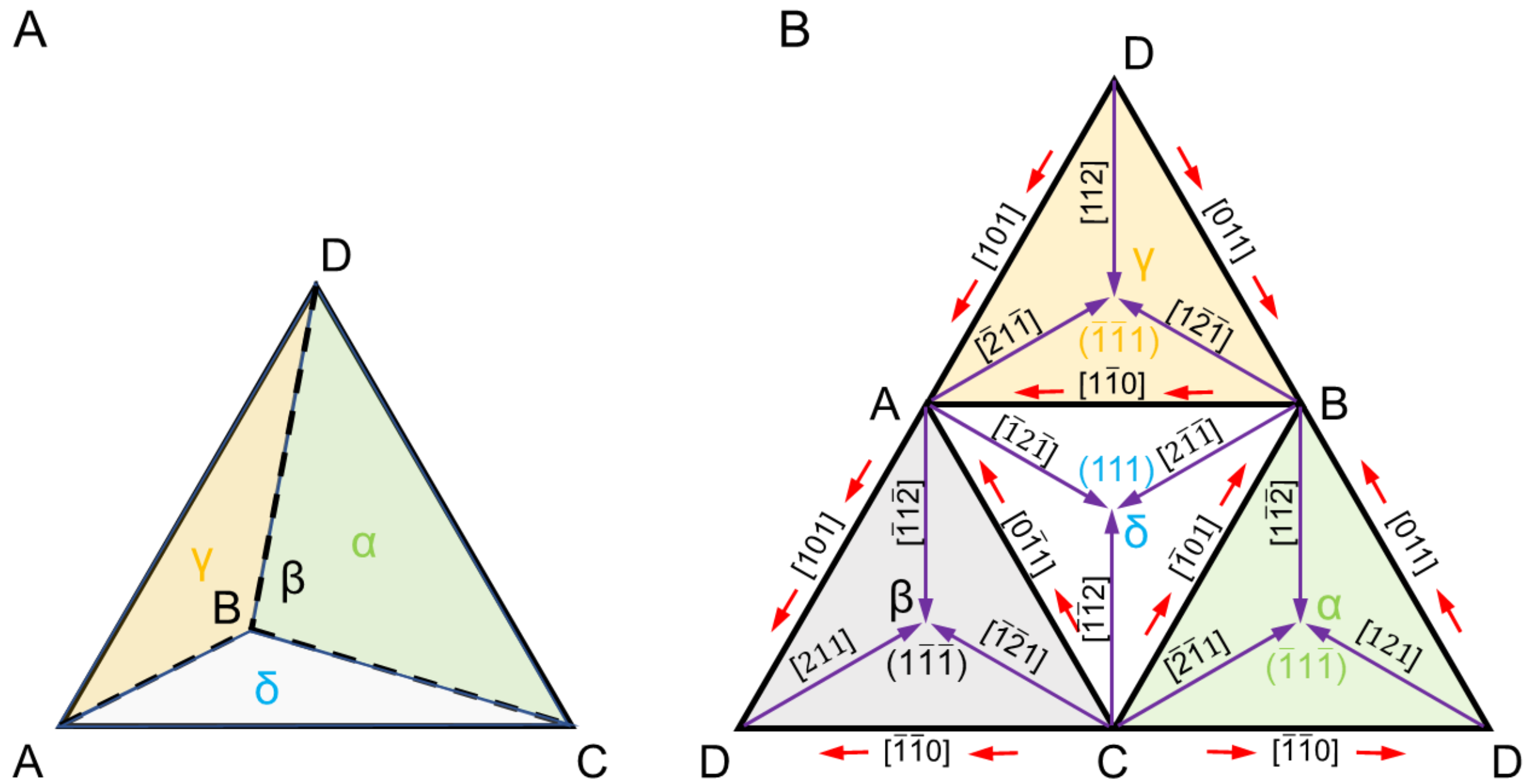

**Figure S10 Thompson tetrahedron in FCC crystals.** (A) Three-dimensional representation of the Thompson tetrahedron. (B) Two-dimensional representation of the Thompson tetrahedron illustrating the possible slip planes and the Burgers vectors of dislocations in an FCC crystal.